\documentclass[rnote]{aa} 
\usepackage{graphicx}
\usepackage{txfonts}
\begin{document}

   \title{Is the anti-correlation between the X-ray variability amplitude and black hole mass of AGNs intrinsic?}


   \author{Yuan Liu
          \inst{1}
          \and
          Shuang Nan Zhang\inst{1}
          }


   \institute{Physics Department and Center for Astrophysics, Tsinghua University, Beijing, 100084,
                China\\
              \email{zhangsn@tsinghua.edu.cn, yuan-liu@mails.tsinghua.edu.cn}
             }

   \date{}


  \abstract
   {}
   {Both the black hole mass and the X-ray luminosity of AGNs have been found to be
   anti-correlated with the normalized excess variance ($\sigma _{\rm rms}^2 $) of the X-ray light curves.
   We investigate which correlation with $\sigma _{\rm rms}^2 $
 is the intrinsic one.}
   {We divide a full sample of 33 AGNs (O' Neill et al. 2005) into two sub-samples.
   The black hole masses of 17 objects in sub-sample 1 were determined by
   the reverberation mapping or the stellar velocity dispersion. The black hole masses
    of the remaining 16 objects were estimated from the relationship between broad
    line region radius and optical luminosity (sub-sample 2). Then partial
     correlation analysis, ordinary least squares regression and K-S tests are performed on the full sample and the sub-samples, respectively.}
   {We find that $\sigma _{\rm rms}^2 $ seems to be intrinsically correlated with the black hole
   mass in the full sample. However, this seems to be caused by including
   the sub-sample 2 into the analysis, which introduces an extra
   correlation between the black hole mass and the luminosity and strengthens any correlation with the black hole mass
    artificially. Therefore,
   the results from the full sample may be misleading. The results from the
    sub-sample 1 show that the correlation between $\sigma _{\rm rms}^2 $ and the X-ray luminosity
    may be the intrinsic one and therefore the anti-correlation between
    $\sigma _{\rm rms}^2 $ and the black hole mass is doubtful.}
   {}


   \authorrunning{Yuan Liu and Shuang Nan Zhang}
   \titlerunning{Anti-correlation between AGN X-ray variability amplitude and luminosity}
   \maketitle
%

\section{Introduction}

   X-ray emission from active galactic nuclei (AGNs) exhibits variability on time scales from
   minutes to days. This indicates that X-rays are likely to be emitted from the inner most
    regions of AGNs and the variability may be related to the important properties of the central
    engine. Lawrence \& Papadakis (1993) utilized long term \emph{EXOSAT} observations to investigate
    the power density spectra of 12 AGNs. They found the power density spectra could be described
     as a power law $P \propto \nu ^{ - \alpha } $ with a mean index $\alpha  = 1.55$, and the amplitude was anti-correlated with the X-ray
      luminosity. Detailed studies of the power density spectra have been performed using
      \emph{RXTE}
      and \emph{XMM-Newton} data. The universal relation between the black hole mass and the "break time"
      in the power density spectra was found both in stellar mass and supermassive black holes
      (e. g. Uttely \& McHardy 2005). A tighter relation was discovered when the bolometric luminosity
      was involved, i.e. $T_B  \propto M^2 /L $ (McHardy et al. 2006). However, due to the
limited observation data,
       the accurate power density spectra are only available for a small number of AGNs. As an alternative,
       the normalized excess variance ($\sigma _{\rm rms}^2 $) can be easily calculated, and it was found that
       $\sigma _{\rm rms}^2 $ is anti-correlated with X-ray luminosity (e.g. Almaini et al. 2000).

       As a result of the progress in determining the black hole masses in AGNs, the relation between
       the variability and black hole mass has also been investigated. Lu \& Yu (2001) found the
       anti-correlation between the black hole mass and $\sigma _{\rm rms}^2 $, and suggested this correlation was an
       intrinsic one, rather than the apparent anti-correlation between the X-ray luminosity and $\sigma _{\rm rms}^2 $.
       Several authors also addressed this problem following Lu \& Yu's work. (e.g. Bian \& Zhao 2003;
       Papadakis 2004; O' Neill et al. 2005). These authors confirmed the anti-correlation between the
        black hole mass and $\sigma _{\rm rms}^2 $, and some models have been subsequently constructed to explain this correlation.

       In this paper we revisit this problem, by studying the sample of O' Neill et al. (2005), which
       includes 33 AGNs and uses nearly the same time scale for all these objects. The black hole
       masses of 17 objects in this sample were determined by the reverberation mapping or the stellar
        velocity dispersion (we denote these objects as sub-sample 1 in the following). The black hole
        masses of the remaining 16 objects were estimated from the relationship between broad line region
        radius and optical luminosity (we denote these objects as sub-sample 2 in the following).
        We find that the optical luminosity (which is used in determining the black hole
        mass for sub-sample 2) has obvious correlation with the X-ray luminosity in 2-10 keV
        band, as shown in Figure 1 (a). The values of the correlation coefficients between optical luminosity and X-ray luminosity for sub-sample 1 and sub-sample 2 are
         0.907 and 0.910, respectively. Even if the datum of NGC 4395 is excluded from sub-sample 2,
         the value of the correlation coefficient is still 0.819. It is well known that there is a strong correlation
         between the optical luminosity and the black hole mass (Kaspi et
        al. 2000). Thus if $\sigma _{\rm rms}^2 $ is intrinsically correlated with one of them, an artificial correlation
        with the other will appear. However, there will be an additional artificial correlation for sub-sample 2 due to
        the utilization of the optical luminosity in determining the black hole mass, whereas the black hole mass of sub-sample 1
        is independently obtained by the reverberation mapping or the stellar velocity dispersion. Therefore, the result from the full sample
        (especially for that from the sub-sample 2) may be misleading. Furthermore, due to the less reliable black hole mass of sub-sample 2, some
         unclear systematic biases may be introduced into the analysis. To find out which
          correlation is intrinsic, we perform the partial correlation analysis to sub-sample 1 and sub-sample 2 separately in \S2.1. The
         ordinary least squares regression results are shown in \S2.2 as another approach
         to this problem. K-S tests are performed in \S2.3 to test whether sub-sample 1 and
         sub-sample 2 are drawn from the same parent population. In \S3 we discuss our results and make conclusions.


\section{Data analysis}

      \subsection{Partial correlation analysis}
The partial correlation analysis is an appropriate method to disentangle the
correlation between variables. The definition of the first order partial correlation
coefficient between variables $x$ and $y$ is (Kendall \& Stuart 1977), $r_{xy.z}  =
\frac{{r_{xy}  - r_{xz} r_{zy} }}{{\sqrt {(1 - r_{xz}^2 )(1 - r_{zy}^2 )} }}$, where
$z$ is the controlled variable and $r_{xy}$ is the correlation coefficient between
variables $x$ and $y$.

We adopt the data of the black hole mass, the X-ray luminosity and
$\sigma _{\rm rms}^2 $ from O' Neill et al. (2005) and present them
in Figure 1 (b) and (c) for clarity. The correlation analysis is
performed on the full sample and the sub-samples, respectively. The
results are shown in Table 1.

   \begin{figure}
   \centering
   \includegraphics[width=6cm]{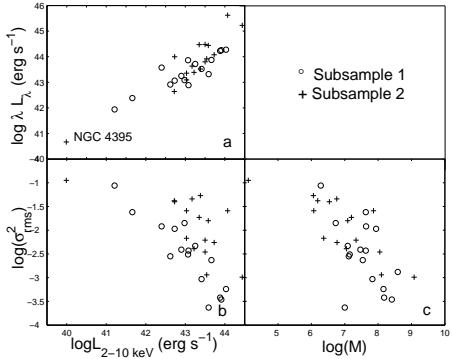}
      \caption{(a): The correlation between X-ray luminosity (2-10 keV) and optical
      luminosity ($\lambda  = 5100\;\mathop {\rm{A}}\limits^ \circ$ ). The point in the left-down corner is NGC 4395, which
      is a dwarf Seyfert galaxy. The value of the correlation coefficient is not significantly
       influenced by whether this point is included or not (see the text for details).
       (b): The data of the normalized excess variances and the X-ray luminosities (2-10 keV).
       (c): The data of the normalized excess variances and black holes masses.
       The values of black hole masses are in units of ${\rm{M}}_ \odot$. The data are obtained
       from O' Neill et al. (2005) and the references therein.}
         \label{FigVibStab}
   \end{figure}

For the full sample, both the black hole mass and the X-ray
luminosity show strong apparent anti-correlations with $\sigma _{\rm
rms}^2 $ (Figure 1 [b] and [c]). However, it appears that after the
black hole mass is controlled, the correlation between the
luminosity and $\sigma _{\rm rms}^2 $ is not significant. On the
contrary, the correlation between the black hole mass and $\sigma
_{\rm rms}^2 $ is still significant after the luminosity is
controlled. These results seem to support Lu \& Yu's suggestion that
the correlation between the black hole mass and $\sigma _{\rm rms}^2
$ is the intrinsic one. However, as discussed in \S1, since any
correlation between the black hole mass may be strengthened
artificially by the sub-sample 2, we should exclude them when
investigating the intrinsic correlation. For sub-sample 1, the
correlation between the black hole mass and $\sigma _{\rm rms}^2 $
disappears when the luminosity is controlled, whereas the
correlation between the luminosity and $\sigma _{\rm rms}^2 $ is
still significant. The results of sub-sample 2 are consistent with
those of the full sample. Both of them indicate $\sigma _{\rm rms}^2
$ is intrinsically correlated with the black hole mass, rather than
the luminosity. However, the analysis of the more reliable
sub-sample 1 shows the contrary results. Due to the limited size of
the present sample, we conclude that the results of sub-sample 2 is
doubtful and maybe the correlation between luminosity and $\sigma
_{\rm rms}^2 $ is the intrinsic one. More robust conclusion will be
deduced when a larger sample is available.

\begin{table*}
\caption{Results of partial correlation analysis.}             
\label{table:1}      
\centering                          
\begin{tabular}{c c c c c}        
\hline                 
 & $r_{L\;\sigma } $& $r_{M\;\sigma } $ & $r_{\;L\;\sigma .M}$ & $r_{M\;\sigma .L} $ \\    
\hline                       
Full sample & -0.636 ($>$99.99\%)   &  -0.697 ($>$99.99\%)   &  -0.277 (87.6\%)  & -0.452 (99.1\%)\\
Sub-sample 1  &   -0.856 ($>$99.99\%) &    -0.560 (98\%)   &  -0.781 ($>$99.99\%)   &  -0.003 (0.8\%)\\
Sub-sample 2  &   -0.605 (98.7\%) &  -0.784 ($>$99.99\%)  &   0.238 (60.8\%) &   -0.653 (99.2\%)\\
\hline                                   
\end{tabular}
\end{table*}

\subsection{Ordinary least squares regression}
To verify the results obtained in \S2.1, we perform the ordinary least square
regression to the full sample, sub-sample 1 and sub-sample 2, respectively. The
regression equation is, $\log (\sigma _{\rm rms}^2 ) = A\log M + B\log L + C.$

The results of the regression for the three samples are summarized
in Table 2. In Figure 2, we show the comparison between the values
of $\sigma _{\rm rms}^2 $ predicted by the results of the
regression and the observed ones.

\begin{table*}

\caption{Results of ordinary least squares regression. The significance of the linear
correlation is obtained by the $F$ statistic. The value of $\chi ^2 $ is calculated
from the regression result to estimate the goodness of the regression. The errors corresponding to 95\% confidence intervals are shown.}             

\label{table:2}      
\centering                          
\begin{tabular}{c c c c c c}        
\hline                 
    & A &  B &  C &  $F$ statistic & $\chi ^2 {\rm{(dof)}}$ \\    
\hline                       
Full sample & -0.38$\pm $0.28 &  -0.22$\pm $0.30 & 10.4$\pm $11.4 & 16.6 ($>$99.99\%)  & 708.6 (30)\\
Sub-sample 1  &   0.00$\pm $0.45 & -0.71$\pm $0.33 & 28.3$\pm $12.2 & 19.2 ($>$99.99\%) &  46.4 (14)\\
Sub-sample 2   &  -0.56$\pm $0.39 & 0.19$\pm $0.45 & -6.06$\pm $17.3 & 11.4 (99.86\%)  &  92.6 (13)\\
\hline                                   

\end{tabular}

\end{table*}

The results of $F$ statistic demonstrate the high significance of
the linear correlation. However, the values of $\chi ^2 $ are still
large, especially for the full sample. We should notice that the
dependence on the black hole mass and the luminosity seems to be
 different for the two sub-samples. For sub-sample 1, $\sigma _{\rm rms}^2 $ appears to depend
 weakly on the black hole mass,
whereas it depends more strongly on the X-ray luminosity. Due to the
small sample, the difference between values of $A$ and $B$ are not
very significant (they are coincidence within the 95\% confidence
interval). However, if the sub-sample 2 is included, the dependence
on the black hole mass is strengthened and the goodness of the
regression decreases dramatically. The value of the total $\chi ^2 $
of the sub-samples is 139 (27); therefore, the probability of the
improvement by chance is only about $10^{ - 9} $ (obtained by
$F$-test). Thus the above results indicate that the sub-sample 1 and
2 are likely to obey different correlation relationships and it is
not appropriate to combine them into one sample.

\subsection{K-S tests}

To investigate whether the two sub-samples are drawn from the same
parent distribution, we first perform the 1D K-S test to the two
sub-samples. The cumulative distribution functions of the two
samples are calculated first. Then the maximum value of the absolute
difference
 between two cumulative distribution functions is used as the statistic to obtain the
 significance of the difference (see details of the K-S test in Press et al. [1992]).
 The significances of the differences are 89\%, 25\% and 96\% for the distributions of
 the black hole mass, the luminosity and $\sigma _{\rm rms}^2 $, respectively (the cumulative distribution
 functions are shown in Figure 3). Clearly except for the X-ray luminosity distribution,
 both the black hole mass and $\sigma _{\rm rms}^2 $ for the two sub-samples are not likely drawn from
 the same parent population. There is no obvious reason accounting for the differences, therefore this result is likely to be due to some unknown selection effects, which
 should be investigated further in the future.

Since it seems visually that the difference between the
distributions of sub-sample 1 and sub-sample 2
  in Figure 1 (b) is more significant than that in Figure 1(c), we perform the 2D K-S test to investigate this problem. The
results of the 2D  K-S test show that the significance of the
difference in Figure 1 (b) is 89.4\%, whereas significance of the
difference in Figure 1 (c) is 97.6\%. This unexpected result is due
to the existence of the point of NGC 4395. After this point is
removed,
  the 2D  K-S test results of Figure 1 (b) and 1 (c) are 96.5\% and 92.1\%, respectively. Although the
  significance of the difference in each figure is high, the visual difference
between two the figures is not significant.

 \begin{figure}
   \centering
   \includegraphics[width=7cm]{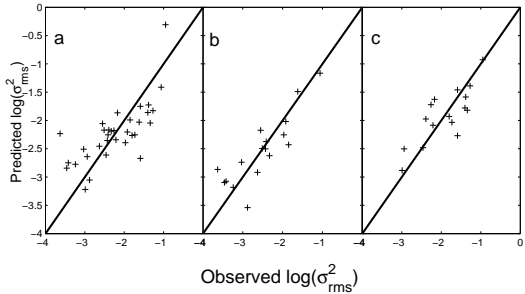}
      \caption{The comparison between the values of $\sigma _{\rm rms}^2 $ predicted by the results of the regression
      and the observed ones. The results from the full sample are shown in (a). (b): The same
      as (a) but for the sub-sample 1. (c): The same as (a) but for the sub-sample 2.}
         \label{FigVibStab}
   \end{figure}

    \begin{figure}
   \centering
   \includegraphics[width=7cm]{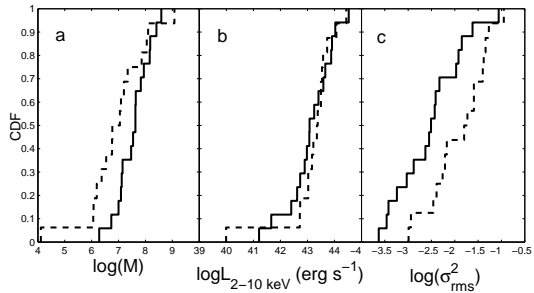}
      \caption{The cumulative distribution functions (CDF) of the black hole mass (a),
      the X-ray luminosity (b) and $\sigma _{\rm rms}^2 $ (c). The solid lines are the data of the sub-sample 1,
      and the dashed lines are the data of the sub-sample 2. The value of black hole mass is in units of ${\rm{M}}_ \odot$.}
         \label{FigVibStab}
   \end{figure}

\section{Discussions and conclusions}

In \S2, we have performed the partial correlation analysis and the
regression on the sample and found that the apparent intrinsic
correlation between $\sigma _{\rm rms}^2 $ and the black hole mass
is likely to be caused by including the sub-sample 2 into the
analysis. Because the black hole masses of AGNs in sub-sample 2 were
estimated from their optical luminosity which in turn is positively
correlated with their X-ray luminosity, an extra correlation between
the black hole mass and X-ray luminosity will be introduced by the
sub-sample 2. If the X-ray luminosity is the primary quantity, then
this will artificially strengthen any correlation with black hole
mass. We therefore should exclude them when investigating the
intrinsic correlation with $\sigma _{\rm rms}^2 $. According to the
results from the sub-sample 1, we conclude that the correlation
between $\sigma _{\rm rms}^2 $
 and the X-ray luminosity may be the
intrinsic one, whereas the apparent correlation between $\sigma _{\rm rms}^2 $
 and the
black hole mass is doubtful. Our K-S tests also suggest that
sub-samples 1 and 2 are not likely drawn from the same parent
population.

As discussed in Lu \& Yu (2001), several mechanisms may be responsible for the
correlation between $\sigma _{\rm rms}^2 $
 and the X-ray
luminosity, such as the hot-spot model, the obscurative variability
and so on. After the apparent correlation between $\sigma _{\rm
rms}^2 $ and the black hole mass was discovered, some models
accounting for this correlation were proposed (e.g. O' Neill et al.
2005, Pessah 2007). However, it needs to be verified whether the
correlation is intrinsic. Although the black hole masses of about
three dozen AGNs have been determined by the reverberation mapping
method, the size of our sample is still limited due to the lack of
long enough and high quality observation data of these objects. More
conclusive results could be obtained when more and higher quality
data become available.

\begin{acknowledgements}
We thank the referee, Andy Lawrence, for valuable comments and suggestions. SNZ
acknowledges partial funding support by Directional Research Project of the Chinese
Academy of Sciences under project No. KJCX2-YW-T03 and by the National Natural Science
Foundation of China under project no. 10521001, 10733010 and 10725313.
\end{acknowledgements}

\end{document}